\documentclass[aps,twocolumn,groupedaddress]{revtex4}

\renewcommand{\vec}[1]{\mbox{\boldmath$#1$}}
\newcommand{\me}{\mathrm{e}}

\newcommand{\dif}{\mathrm{d}}

\usepackage{graphicx}

\begin{document}
\bibliographystyle{apsrev}
\title{Self Consistent Molecular  Field Theory for Packing in Classical
Liquids}
\author{Lawrence  R.  Pratt}
\affiliation{Theoretical Division, Los Alamos National Laboratory, Los
Alamos, NM 87545, USA}

\author{Henry S. Ashbaugh}

\affiliation{Theoretical Division, Los Alamos National Laboratory, Los
Alamos, NM 87545, USA}

\date{\today}

\begin{abstract}
Building on a quasi-chemical  formulation of solution theory, this paper
proposes a self consistent molecular field theory for packing problems
in classical liquids, and tests the theoretical predictions for the
excess chemical potential of the hard sphere fluid.  Results are given
for the self consistent  molecular fields  obtained, and for the
probabilities of occupancy of a molecular observation volume.  For this
system, the excess chemical potential predicted  is as accurate as the
most accurate prior theories, particularly the scaled particle
(Percus-Yevick compressibility) theory.  It is argued that the present 
approach is particularly simple, and should provide a basis for a
molecular-scale description of more complex solutions.
\end{abstract}

\maketitle

\section{Introduction}
The disordered packing of  molecules at liquid densities is a  primary
and difficult problem in the theory  of liquids \cite{Widom:67,WCA}.
This problem  is typically addressed first by consideration of model
intermolecular interactions of hard-core type, interactions that rigidly
exclude  molecular overlap.   For those systems, a quantity of primary
interest is then Boltzmann's \emph{available phase space}
\cite{Stell:MM:85} from which follows the thermodynamic excess chemical
potential discussed here. Sophisticated  theories, even if esoteric, are
available \cite{Reiss:59,WERTHEIMMS:EXASOP,THIELEE:EQUOSF,Reiss:77} for
the equation of state of the hard sphere fluid.  In conjunction with
simulation results, adaptations of those theories provide empirically
exact results for the hard sphere system \cite{Hansen}.  Recent
theoretical activity \cite{Crooks:PRE:97,Weeks:JCP03} on the hard sphere
fluid emphasizes that physical clarity is an important quality of
theories that might be transplanted to describe more realistic solution
models.  The physical content of available models of packing of more
realistically shaped molecules is conceptually similar to theories of
the hard sphere fluid, but the resultant theories are naturally more
complicated than for hard spheres;
Refs.~\cite{BOUBLIKT:HARCBE,Reiss:81,WERTHEIMMS:FLUOHC,LabikS:Comscp,%
LEELL:APDA,LUEL:ANASOT,MehtaSD:Equsav,MehtaSD:GenFth,BarrioC:Equshc,%
LargoJ:Equsff,Ben-AmotzD:Cavfed,JafferKM:Thenpt,BoublikT:Selesh,%
BjorlingM:OnaFa,Crooks:PRE:97} give examples of that ongoing activity.

Recent developments of a quasi-chemical approach to the theory of
molecular solutions \cite{Paulaitis:02} have brought a new set of
concepts to bear on these problems \cite{PrattLR:Quatst}; these
developments suggest  theories with clear physical content and a  clear
path for progressive improvement.   This paper pursues these
developments further, proposing and testing a self consistent molecular
field theory for molecular scale packing.  More important than the
specific packing problem considered here, these self consistent
molecular field ideas will be carried forward to develop quasi-chemical
treatments of realistic solutions \cite{Paulaitis:02}.   

\begin{figure}[h!]
\begin{center}
\leavevmode
\includegraphics[scale=0.75]{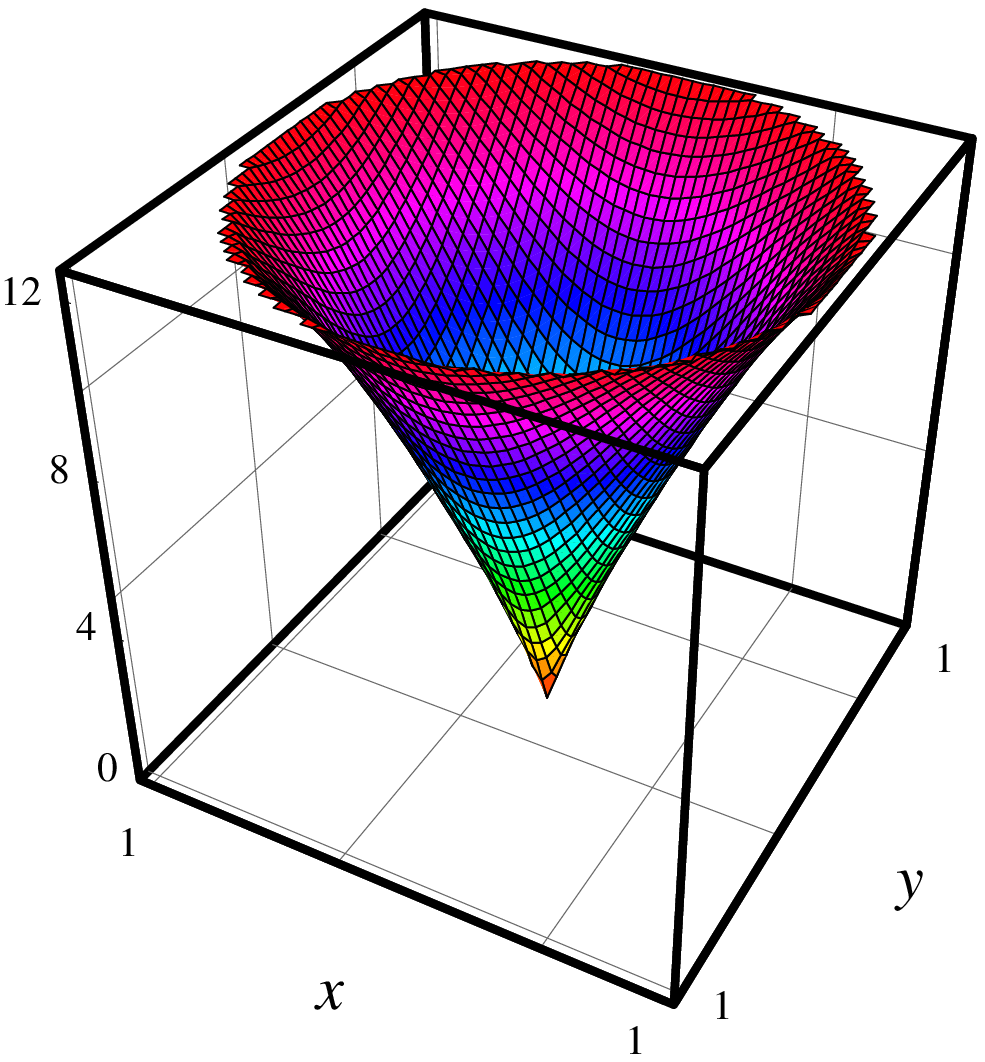}
\includegraphics[scale=0.6]{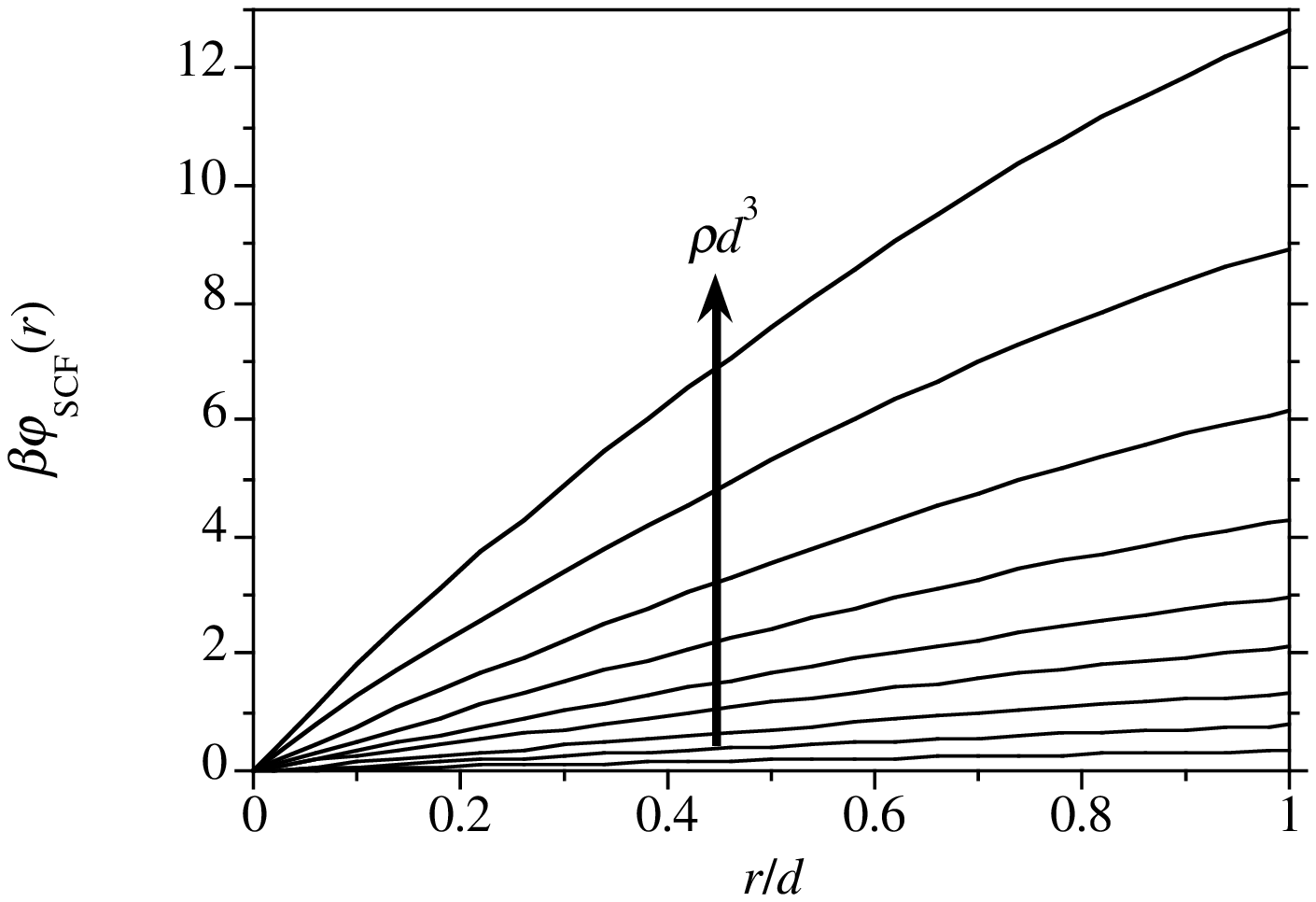} 
\end{center}
\caption{The self consistent molecular field
$\beta\varphi_{SCF}(\vec{r})$ for $d$-diameter hard spheres a spherical
observation volume of radius $d$.  $r/d$ = 0 is the center of the
observation volume, and $r/d$ = 1 is the surface.  The curves on the
bottom panel correspond, from bottom  to top, to reduced densities $\rho
d^3$  = 0.1, \ldots, 0.9, in increments of 0.1.  The upper panel depicts
$\beta\varphi_{SCF}$ for $\rho d^3$  = 0.9, on a plane through the
center of the observation  sphere.}
\label{fig1}
\end{figure}

\section{Theory}

For economy of the presentation, we specifically discuss the one
component hard sphere fluid.   The quasi-chemical theory is built upon
the relations \cite{PrattLR:Quatst,Paulaitis:02}
\begin{eqnarray}
p_n(R)={{K_n(R)\rho ^n} \over {1+\sum\limits_{m\ge 1} {K_m(R)\rho ^m}}}
\label{pn}
\end{eqnarray}
where $p_n(R)$ are probabilities for observing $n$ sphere centers in an
observation sphere of radius $R$ in the liquid,  and the $K_n(R)$ are
well defined equilibrium ratios of concentrations of hard sphere
$n$-complexes with that observation sphere with $K_0\equiv$ 1.  The
quantities $K_n(R)$ describe  occupancy transformations fully involving
the solution neighborhood of the observation volume.  Except in the
limit of low density, these coefficients are  known only approximately. 
Therefore, physically motivated approximations are  required to proceed
to specific quantitative predictions.

Our previous study  of this problem \cite{PrattLR:Quatst} identified a
primitive quasi-chemical approximation in which
\begin{eqnarray}
K_n( R)  & \approx &   {\zeta^n \over {n!}}\int\limits_v d\vec{r}_1\ldots
\int\limits_v d\vec{r}_n
\me^{
- \sum\limits_{i>j=1}^n \beta u(\vec{r}_{ij})}~.
\label{pqca}
\end{eqnarray}
Here $v=4\pi R^3/3$ is the volume of the observation sphere,
$\beta^{-1}$ = $kT$, $u(\vec{r}_{ij})$ is the interaction between
molecules $i$ and $j$ (the hard sphere interaction in the present case),
and $\zeta$ is a Lagrange multiplier used to achieve consistency between
the known bulk density, $\rho$ , and the average density in the
observation volume. Because of the explicit factors of $\rho$ in
Eq.~\ref{pn},  $\zeta$ will approach the thermodynamic excess activity,
$\ln \zeta = \beta\mu^{ex}$ with $\mu^{ex}$ the excess chemical
potential of Gibbs. The integrals of Eq.~\ref{pqca} are few-body
integrals that can be estimated by Monte Carlo methods
\cite{PrattLR:Quatst}.  A natural extension of this idea is to
approximate $K_n(R)$ on the basis of $n$-molecule configurational
integrals that give the low-density limiting quantity, but with
inclusion of a molecular field $\beta\varphi_{SCF} (\vec{r})$ as
\begin{eqnarray}K_n( R)  & \approx &   {\zeta^n \over {n!}}\int\limits_v d\vec{r}_1\ldots
\int\limits_v d\vec{r}_n
\me^{-\sum\limits_{i=1}^n \beta\varphi_{SCF}(\vec{r}_i)
- \sum\limits_{i>j=1}^n \beta u(\vec{r}_{ij})}
\nonumber \\   & \equiv &  K_n^{(0)}(R;\beta\varphi_{SCF} )~.
\label{scf}
\end{eqnarray} 
We will adopt the convention that the molecular field
$\beta\varphi_{SCF} (\vec{r})$ be zero at the center of the observation
volume, and an  additive constant be absorbed in the Lagrange
multipliers of the $K_n^{(0)}(R;\beta\varphi_{SCF} )$. The molecular
field $\beta\varphi_{SCF} (\vec{r})$,  together with the Lagrange
multiplier,  may be made consistent with the information that the
prescribed density of the liquid is uniform within the observation
volume. The density profile for the $n$-molecule case is \cite{funcdiff}
\begin{eqnarray}
\rho _n(\vec{r} )  =    - {\delta \ln K_n^{(0)}(R;\beta\varphi_{SCF} ) \over \delta \beta\varphi_{SCF} (\vec{r})}
\label{rho-n}
\end{eqnarray}
inside the observation volume.  Averaging of these profiles with respect
to the possible occupancies predicts the observed density.  The
consistency sought is  then uniformity of the density,
\begin{eqnarray}
- \sum_m p_m\- {\delta \ln K_m^{(0)}(R;\beta\varphi_{SCF} ) \over \delta
\beta\varphi_{SCF} (\vec{r})}     =  -  {\delta \ln p_0 \over \delta
\beta\varphi_{SCF} (\vec{r})}  \nonumber \\ = {\delta \beta\mu^{ex}
\over \delta \beta\varphi_{SCF} (\vec{r})}   =  \rho ~,
 \label{self-con}
\end{eqnarray}
for $\vec{r}$ inside the observation volume.  $p_0$ is defined by
Eq.~\ref{pn}, and in Eq.~\ref{self-con} we have noted that, for
hard-core solutes, the interaction contribution to the chemical is
$\beta\mu^{ex}$ = -$\ln p_0$ \cite{PrattLR:Quatst,Paulaitis:02}.

\begin{figure}[h!]
\includegraphics[scale=0.6]{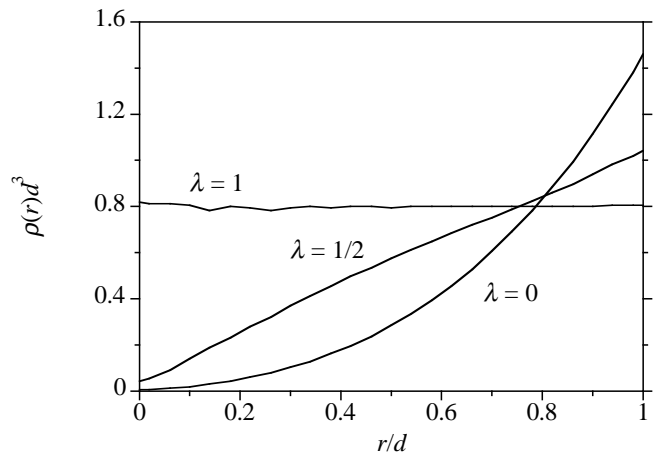}
\caption{Example dependence of the density profile on scaled molecular
field $\lambda\beta\varphi_{SCF} (\vec{r})$; $\rho d^3$ = 0.8.}
\label{fig2}
\end{figure}

\begin{figure}[h!]
\includegraphics[scale=0.6]{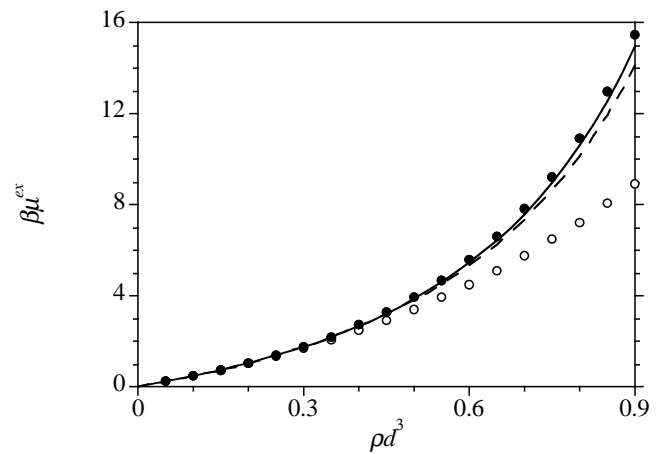}
\caption{Excess chemical potential of the hard sphere fluid as a
function of density.  The open and filled circles correspond to  the
predictions of the primitive quasi-chemical theory and the present self
consistent molecular field theory, respectively.  The solid and dashed
lines are the scaled particle (Percus-Yevick compressibility) theory and
the Carnahan-Starling equation of state, respectively.}
\label{fig3}
\end{figure}

\begin{figure}[h!]
%\begin{center}
%\leavevmode
\includegraphics[scale=0.6]{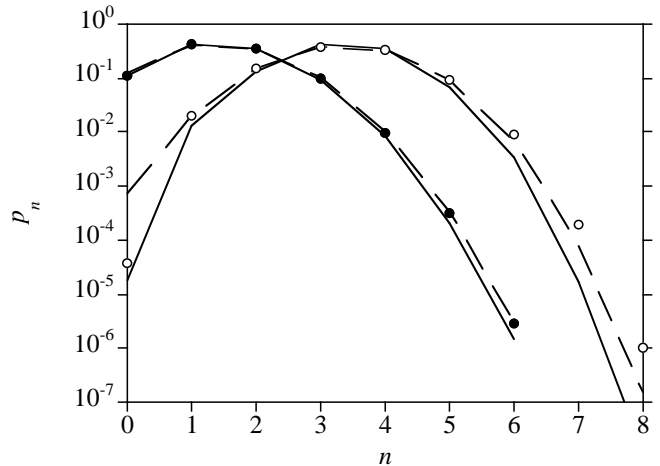}
%\end{center}
\caption{Distributions $p_n$ with $R$=$d$ for densities of $\rho d^3$ =
0.35 (filled circles) and 0.8 (open circles).  The dashed lines are the
primitive quasi-chemical theory  of  Ref.~\cite{PrattLR:Quatst}, and the
solid lines correspond to the present  SCF theory.  Note the marked
\emph{break-away} of the $n$=0 point from the  primitive quasi-chemical
curve, observed before \cite{PrattLR:Quatst}.  The errors on the high
$n$ side of these distributions might reflect the fact that the present
SCF theory doesn't  explicitly  treat  pair correlations.  Those
correlations enter only through the integrals
$K_n^{(0)}(R;\beta\varphi_{SCF} )$.}
\label{fig4}
\end{figure}

\begin{figure}[h!]
%\begin{center}
%\leavevmode
\includegraphics[scale=0.55]{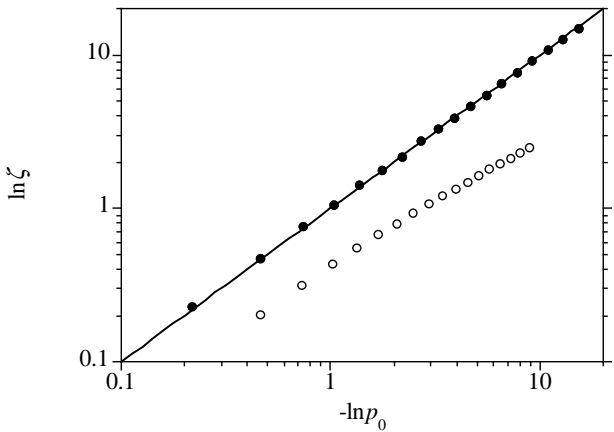}
%\end{center}
\caption{Comparison of $\ln\zeta$ (with $\zeta$ the Lagrange multiplier 
or excess activity) against computed excess chemical potential,
$\beta\mu^{ex}$ = -$\ln p_0$, demonstrating the thermodynamic
consistency of these quasi-chemical theories.  The open circles are the
primitive quasi-chemical theory (Eq.~\ref{pqca}), and the filled circles
are the  present self consistent molecular field  theory.}
\label{fig5}
\end{figure}

Examples of the results following from these ideas are shown in
Figs.~\ref{fig1}-\ref{fig5}.  These results were obtained from a two
step iterative procedure from a starting guess $\beta\varphi_{SCF}
(\vec{r})$ = 0 and the probabilities $p_n$ of the primitive
quasi-chemical theory \cite{PrattLR:Quatst}.    With the current
approximate results, we performed  Monte Carlo calculations to estimate
the  densities for each occupancy, and on that basis the average density
implied by the current field.  We then  updated  the molecular field
according to
\begin{eqnarray}
\left\lbrack \beta\varphi(\vec{r})_{SCF} \right\rbrack_{new} =
\left\lbrack \beta\varphi(\vec{r})_{SCF} \right\rbrack_{old} + f \ln 
\left\lbrack \frac{\rho(r)}{\rho}\right\rbrack~,
\end{eqnarray}
where $f$ is a constant less than one that ensures stable convergence of
the molecular field; a value of 0.2 was  found to work here. 
Convergence is obtained in 20 iterations of this procedure, typically.
Using the  field obtained above, we then updated the occupancies,
reevaluating the $K_n^{(0)}(R;\beta\varphi_{SCF})$    by performing
additional few-body simulations to evaluate the work associated with
turning on the molecular field using thermodynamic integration
\begin{eqnarray}
\frac{K_n^{(0)}(R;\beta\varphi_{SCF} )}{ K_n^{(0)}(R;0)}  =  \me^{ -
\int_0^1 \left \langle \sum_{j=1}^n\beta\varphi_{SCF}
(\vec{r}_j)\right\rangle_\lambda\dif\lambda } \nonumber \\
\end{eqnarray}
where $\lambda$ is a coupling parameter, and $\left
\langle\ldots\right\rangle_\lambda$ indicates averaging over
configurations generated under the influence of the molecular field
scaled as $\lambda\beta\varphi_{SCF} (\vec{r})$. Using these
recalculated $K_n^{(0)}(R;\beta\varphi_{SCF} )$, we  generated a new set
of $p_n$, tested for convergence, and so on. This process was found to
converge within two steps even at the highest densities considered.  We
attribute the observed convergence to the fact that the starting point,
the primitive quasi-chemical theory, is accurate for the probable
occupancies. The molecular fields obtained using this method were found
to converge stably  with little difficulty.

Fig.~\ref{fig1} shows the self consistent molecular fields obtained
using the procedure described above up to fluid densities of $\rho d^3$
= 0.9, just below the hard sphere freezing transition.
$\beta\varphi_{SCF} (\vec{r})$ is a monotonically increasing function of
radial position from the center of the stencil volume to its boundary.
This reflects the fact that in the absence of the molecular field the
hard sphere particles tend to build up on the surface of the stencil
volume  to minimize their interactions with the other particles
(Fig.~\ref{fig2}). The molecular field makes the boundary repulsive,
depletes the surface density, and homogenizes the density within the
volume. The magnitude of this repulsive field increases with increasing
fluid density.

The predicted hard sphere chemical potentials as a function of density
using the primitive and self consistent molecular field quasi-chemical
theories are compared to the  chemical potential from the
Carnahan-Starling equation in Fig.~\ref{fig3}. The primitive theory
works well up to $\rho d^3\approx$ 0.35, roughly the critical density
for  Ar and the density region suggested to mark qualitative packing
changes in the hard sphere fluid \cite{GIAQUINTAPV:STRROT};  at higher
densities the primitive quasi-chemical theory systematically
under-predicts the hard sphere chemical potential. The present self
consistent molecular field theory significantly improves the agreement
with the Carnahan-Starling equation over the entire density range. Above
densities of $\rho d^3\approx$ 0.6, the self consistent molecular field
theory begins to over-predict the hard sphere chemical potential, though
the absolute value of the error is in marked improvement over the
primitive theory. We note that the self consistent molecular field
theory is in closer agreement with the scaled particle (or Percus-Yevick
compressibility) theory for the chemical potential.   
Fig.~\ref{fig4} shows that the most important deficiencies of the 
primitive quasi-chemical theory are corrected by the self-consistent
molecular field theory.  Note that the self-consistent molecular field
theory captures the \emph{break-away} at high density of $\ln p_0$ from
the primitive quasi-chemical prediction.

In addition to achieving a uniform density across the observation
volume, the self consistent molecular field also nearly achieves
thermodynamic consistency for the chemical potential.  With the choice
of an additive constant which makes $\beta\varphi_{SCF} (\vec{r})$ zero
in the deepest interior of the observation volume,  $\ln\zeta$ should
approach the excess activity of the solvent in the limit of a large
observation volume.  We expect  on a physical basis that
$\beta\varphi_{SCF} (\vec{r})$ describes an interaction between the
interior and the  exterior of the observation volume across the
intervening surface.  Particularly in the present case of  short ranged
interactions, we expect spatial variation of $\beta\varphi_{SCF}
(\vec{r})$ to be confined to a surface region.  Though a stencil volume
of radius $R=d$ is evidently  not large enough to observe that bulk
behavior (Fig.~\ref{fig1}), for  that $R=d$ case we can compare the
computed excess chemical potential with the solvent activity.  
Fig.~\ref{fig5} compares -$\ln p_0$ and  $\ln \zeta$  as determined by
the primitive and self consistent molecular field quasi-chemical
theories. While the activity evaluated within the primitive theory
significantly under-predicts $p_0$, with the self consistent molecular
field theory  $\ln \zeta$ and  -$\ln p_0$ agree nearly perfectly.  At
the highest densities, there is a slight disparity between these two
quantities, and the calculated values for $\ln \zeta$ are in better
agreement with the empirically known $\beta \mu ^{ex}$ for the hard
sphere fluid.

\section{Variation with Cavity Size}
The related quantity
\begin{eqnarray}
  4\pi \rho R^2G(R)  & = & -  \frac{\dif\ln  p_0}{\dif R} \end{eqnarray}
  is of special interest in the theory of the hard sphere fluid, and of
  solubility more generally \cite{Stillinger:73,Reiss:77,HSA:03}.  In
  the present quasi-chemical approximation, this is \begin{eqnarray}
  4\pi \rho R^2G(R)   \approx & \sum_m p_m\left( \dif \ln
  K_m(R;\beta\varphi_{SCF} ) / \dif  R\right)~.
\end{eqnarray}
To analyze the required derivative, we consider that the radius $R$ is
defined in the first place by a bare field $\beta\varphi_{0}$ that is
zero (0) inside the observation volume and $\infty$ outside.  Then the
full field encountered with the integrals Eq.~\ref{scf} is
$\beta\varphi$  = $\beta\varphi_{0}$ + $\beta\varphi_{SCF}$. The result
now corresponding to Eq.~\ref{rho-n} is
\begin{eqnarray}
{{\dif \ln K_m(R;\beta\varphi_{SCF}   )} \over {\dif  R}} = - \int_v
\rho _m(  \vec{r};\beta\varphi_{SCF}  ) {{\partial \beta\varphi
(\vec{r})} \over {\partial R}}\dif^3 r~. \nonumber \\
\label{chain-rule-3}
\end{eqnarray}
Upon  separating the contribution from $\beta\varphi_{0}$ and performing
the population averaging, this produces the simple relation
\begin{eqnarray}
  4\pi \rho R^2G(R)=  4\pi R^2\rho  - \int_v  {{\partial
  \beta\varphi_{SCF} (\vec{r})} \over {\partial R}}\rho  \dif^3 r.
\label{chain-rule-4}
\end{eqnarray}
With this first population averaging, we emphasize that
$\beta\varphi_{SCF} (\vec{r})$ doesn't depend on the occupancy index
$m$. The radius derivative $\partial \beta\varphi(\vec{r}) /\partial R$
of the full field can be described by a simple formal relation.  The
relation
\begin{eqnarray}
\frac{\delta \rho_m(\vec{r})}{\delta \beta\varphi(\vec{r}')} =
-\left\langle \delta \rho_m(\vec{r})\delta\rho_m(\vec{r}')\right\rangle~
\label{n-sus}
\end{eqnarray}
follows from Eq.~\ref{rho-n} for each occupancy.  Performing the 
population averaging at this stage, we write
\begin{eqnarray}
-\frac{\delta \rho(\vec{r})}{\delta \beta\varphi (\vec{r}')} =
\left\langle \delta \rho(\vec{r})\delta\rho(\vec{r}')\right\rangle\equiv
\chi(\vec{r},\vec{r}')
\label{susceptibility}
\end{eqnarray}
and 
\begin{eqnarray}
 - \delta \beta\varphi(\vec{r}) = \int \chi^{-1}(\vec{r},\vec{r}')\delta
 \rho(\vec{r}') \dif^3  r' ~.
 \label{increment}
\end{eqnarray}
Averaging of the functional derivative Eq.~\ref{n-sus} \emph{before}
composing Eq.~\ref{increment} is suggestive of the RPA concept of
exploiting an  average potential in a linear response function. To use
Eq.~\ref{increment},  consider the  density change $\delta
\rho(\vec{r}')$ corresponding to dematerialization of the uniform
density in a thin  shell $\left(R-\Delta R,R \right)$.
\begin{eqnarray}
- \frac{\partial\beta\varphi(\vec{r})}{\partial R} =  R^2 \rho
\int_{\left|\vec{r}'\right|=R_{-}} \chi^{-1}(\vec{r},\vec{r}') \dif^2 
\Omega' ~,
\end{eqnarray}
where the latter integral is over solid angles covering the surface  of
the ball.  We introduce now $c(\vec{r},\vec{r}')$,  the Ornstein-Zernike
direct correlation function defined by $ \chi^{-1}(\vec{r},\vec{r}')$ =
$\delta(\vec{r}-\vec{r}')/\rho(\vec{r})-c(\vec{r},\vec{r}')$. We finally
obtain
\begin{eqnarray}
G(R) & = & 1 -  \int_v   c(\vec{r},\vec{r}'=\hat z R)\rho  \dif^3 r 
\label{finally}
\end{eqnarray}
within the present approximation.  In the indicated integral the
$\vec{r}'$ coordinate is pinned to the sphere surface, and $\vec{r}$
integration is over the interior of the sphere because of
Eq.~\ref{chain-rule-3}.     The function $c(\vec{r},\vec{r}')$ is the OZ
direct correlation function in the  field $\beta\varphi$ including the
self consistent molecular field, thus for the case of a uniform density
enclosed in a  sphere of radius $R$ with  no materal outside.

It is obvious that Eq.~\ref{finally} gives the corrrect answer for the
case that  the solvent atoms have no interaction with one another
($c(\vec{r},\vec{r}')$= 0), and for the same reason this formula is
obviously correct in the limit of zero density.  That limiting result
gives the second virial coefficient theory for $\ln p_0$.  At the
initial order in the density
\begin{eqnarray}
c(\vec{r},\vec{r}') = \exp[-\beta u(\vec{r},\vec{r}')] - 1 + O(\rho).
\label{c}
\end{eqnarray}
This relation in the approximate Eq.~\ref{finally} leads to the correct
contribution of  next order in the density for $G(R)$, corresponding the
third virial contribution to $\ln p_0$.

Exact results are also available in the  case that the observation
sphere is sufficiently small,   $R\le\frac{d}{2}$.  Then $p_0$ = 1-$4\pi
\rho R^3/3$, $\beta\varphi_{SCF}(\vec{r})$=$\ln\left(1-4\pi\rho
R^3/3\right)$ (spatially uniform in $0\le r\le R$, so in the formulation
above this  would be reflected solely in the Lagrange multipliers). 
Direct calculation gives  $\chi(\vec{r},\vec{r}')$ =
$\rho\delta(\vec{r}-\vec{r}')-\rho^2$, and $\chi^{-1}(\vec{r},\vec{r}')$
= $\rho^{-1}\delta(\vec{r},\vec{r}')$ + $\frac{1}{1-4\pi \rho R^3/3}$. 
Using these results in Eq.~\ref{finally} gives the known answer,
$G(R)$=$\frac{1}{1-4\pi \rho R^3/3}$.  Tests of other current theories
in this regime have been given by \cite{Weeks:JCP03}.

\section{Concluding Discussion}

The physical content of the present self consistent molecular field
theory is simple and  clear,  and this theory is as accurate as the most
accurate prior theories, particularly the scaled particle (Percus-Yevick
compressibility) theory, for the thermodynamics of the hard sphere
fluid.  The conclusion is that careful attention to the near
neighborhood of a distinguished solute in such a liquid, with a self
consistent molecular field describing the influence  of the more distant
neighborhood,  provides an accurate description of packing in dense
liquids.  Though distinct, the \emph{hydrostatic linear response} theory
\cite{Weeks:JCP03} leads to a similar conclusion that good theories of 
these phenomena can be extremely local.  The present results address
contributions essential to quasi-chemical descriptions  of solvation in
more realistic cases, as has been discussed on a conceptual basis
recently \cite{Paulaitis:02}.

The present results provide a definite,  and organized basis for
theoretical study  of subsequent solvation  phenomena.  For example,
consider inclusion of attractive interactions between solvent molecule
spheres, interactions secondary to  the repulsive interactions.  The
simple estimate $c(r) \sim -\beta u(r)$ for distances not too small,  is
consistent with Eq.~\ref{c}.  But when $u(r)$ at those distances
describes attractive interactions, Eq.~\ref{finally} then predicts that
these attractive interactions reduce the magnitude of $G(R)$.  This is a
behavior that has been much discussed over recent years in the  context
of theories of inert gas solubility in liquid water
\cite{PrattLR:Quatst,Pratt:2002,Paulaitis:02,HSA:03}.

A related but distinct issue is how these packing questions are affected
by multiphasic behavior of the  solution, particularly the possibility
of \emph{ drying} \cite{LCW,Weeks:2002,ChandlerD:HydTfw} or preferential
absorption \cite{DiazMD:Evichc} in biophysical aqueous applications.  In
such cases, it is attractive to speculate that the self consistent
molecular field $\beta\varphi_{SCF}$ should reflect those multiphase
possibilities just as it can in pedagogical treatments of non-molecular
models of phase transitions \cite{Ma}.

\bigskip
\section*{Acknowledgements}

We thank for Dilip Asthagiri and Michael E. Paulaitis for discussions
and comments  on a preliminary draft of  this paper.  This work was
supported by the US Department of Energy, contract W-7405-ENG-36, under
the LDRD program at Los Alamos. LA-UR-03-3111.

% If in two-column mode, this environment will change to single-column
% format so that long equations can be displayed. Use
% sparingly.
%\begin{widetext}
% put long equation here
%\end{widetext}

% figures should be put into the text as floats.
% Use the graphics or graphicx packages (distributed with LaTeX2e).
% See the LaTeX Graphics Companion by Michel Goosens, Sebastian Rahtz,
% and Frank Mittelbach for instance.
%
% Here is an example of the general form of a figure:
% Fill in the caption in the braces of the \caption{} command. Put the label
% that you will use with \ref{} command in the braces of the \label{} command.
%
% \begin{figure}
% \includegraphics{}%
% \caption{}
% \label{}
% \end{figure}

% tables follow here or maybe be put in the text
%
% Here is an example of the general form of a table:
% Fill in the caption in the braces of the \caption{} command. Put the label
% that you will use with \ref{} command in the braces of the \label{} command.
% Insert the column specifiers (l, r, c, d, etc.) in the empty braces of the
% \begin{tabular}{} command.
% The ruledtabular enviroment adds doubled rules to table and sets a
% nice set of default table settings.
% Use the table* environment to get a full-width table in two-column
% \begin{table}
% \caption{}
% \label{}
% \begin{ruledtabular}
% \begin{tabular}{}
% \end{tabular}
% \end{ruledtabular}
% \end{table}

% If you have acknowledgments, this puts in the proper section head.

% Create the reference section using BibTeX:
%\bibliography{book}

\end{document}